\documentclass[twocolumn,english,5p,sort&compress,times]{elsarticle}
\usepackage[T1]{fontenc}
\usepackage[latin9]{inputenc}
\usepackage{amsmath}
\usepackage{graphicx}
\usepackage{setspace}
\usepackage{amssymb}
\makeatletter

\providecommand{\tabularnewline}{\\}

\journal{Icarus}

\usepackage{hyperref}
\makeatother\usepackage{lineno}\usepackage{babel}\usepackage{multirow}\usepackage{multicol}
\usepackage{amsmath}
\makeatother

\usepackage{babel}

\makeatother

\usepackage{babel}

\begin{document}

\title{Discovery Prospects for a Supernova Signature of Biogenic Origin}

\author[rvt]{S. Bishop\fnref{fn1}}

\ead{shawn.bishop@ph.tum.de}

\author[coauth]{R. Egli}

\fntext[fn1]{Corresponding author. Fax: +49 089 289 12435}

\address[rvt]{Physik Department E12, Technische Universität München, D-85748 Garching,
Germany}

\address[coauth]{Department of Earth and Environmental Sciences, Ludwig-Maximilians
University, Theresienstrasse 41, 80333 Munich, Germany}
\begin{abstract}
Approximately 2.8~Myr before the present our planet was subjected
to the debris of a supernova explosion. The terrestrial proxy for
this event was the discovery of \emph{live} atoms of $^{60}\mbox{Fe}$
in a deep-sea ferromanganese crust. The signature for this supernova
event should also reside in magnetite $(\mbox{Fe}_{3}\mbox{O}_{4})$
microfossils produced by magnetotactic bacteria extant at the time
of the Earth-supernova interaction, provided the bacteria preferentially
uptake iron from fine-grained iron oxides and ferric hydroxides. Using
estimates for the terrestrial supernova $^{60}\mbox{Fe}$ flux, combined
with our empirically derived microfossil concentrations in a deep-sea
drill core, we deduce a conservative estimate of the $^{60}\mbox{Fe}$
fraction as $^{60}\mbox{Fe}/\mbox{Fe}\approx3.6\times10^{-15}$. This
value sits comfortably within the sensitivity limit of present accelerator
mass spectrometry capabilities. The implication is that a biogenic
signature of this cosmic event is detectable in the Earth's fossil
record.\end{abstract}
\maketitle

\section{Introduction}

Cosmic production sites for $^{60}\mbox{Fe}$ are massive stars which
produce $^{60}\text{Fe}$ during their quiescent helium- and carbon-shell
burning phases~\citep{limongi03,limongi06}, followed by subsequent
shock heating of their helium and carbon shells during the core collapse
stage of their evolution~\citep{limongi03,limongi06,rauscher02}.
Carbon deflagration Type Ia supernovae are another potential site
for $^{60}\mbox{Fe}$ production~\citep{woosley97,nomoto84}. Both
sites can produce up to $10^{-4}\, M_{\odot}$ of $^{60}\mbox{Fe}$.
Radioactive $^{60}\mbox{Fe}$, with a newly revised half-life of 2.62~Myr~\citep{rugel09},
has a $\beta$-decay scheme~\citep{rugel09} that gives rise to two
gamma-rays, from the decay of excited states in $^{60}\mbox{Ni}$,
with energies, $E_{\gamma_{1}}=1173$~keV and $E_{\gamma_{2}}=1332$~keV.
It is an important radionuclide for tracing active nucleosynthesis
within our galaxy because its half-life is much shorter than stellar
lifetimes, yet also long as compared to the expansion time scales
of supernovae ejecta thereby allowing sufficient amounts of it to
survive the opaque ejecta phase and be susceptible to observation
with gamma-ray astronomy. Recent gamma-ray astronomy studies have
detected these gamma-rays within the inner radiant portion of the
galactic plane~\citep{wang07}, known to be a site of massive stars.

Using the accelerator mass spectrometry (AMS) facility~\citep{knie00,knie97}
of the Maier-Leibnitz-Laboratory, operated by the Munich universities,
a concentration spike of \emph{live} $^{60}\mbox{Fe}$ atoms was discovered~\citep{knie99,knie04}
in a Pacific Ocean deep-sea ferromanganese crust (FMC). The measured
$^{60}\mbox{Fe}$ atom concentration was $^{60}\mbox{Fe}/\mbox{Fe}\approx2\times10^{-15}$.
From these data, it was concluded that our planet has been exposed
to the debris of a supernova (SN) $\approx2.8$~Myr before the present~\citep{knie04}.
This interpretation was subsequently challenged~\citep{basu07} on
the basis of highly variable FMC $^{3}\mbox{He}$ concentrations within
the SN epoch. The variation in the $^{3}\mbox{He}$ concentration
was attributed to the sampling of micrometeorites trapped in the FMC.
The challenge thus made was that the discovered $^{60}\mbox{\mbox{Fe}}$
was of cosmogenic origin, produced by galactic cosmic ray spallation
reactions on nickel within the micrometeorites. This challenge was
subsequently refuted on quantitative grounds using the nickel and
iron abundances of the FMC itself~\citep{fitoussi08} and finding
the identical $^{60}\mbox{Fe/\mbox{Fe}}$ signal in a different region
of the original FMC~\citep{fitoussi08} and, additionally, finding
this same signal in a different FMC (29DR-32) obtained from the northern
Pacific Ocean~\citep{korschinek_fmc}. We are, ourselves, thus inclined
to discount the micrometeorite hypothesis and consider a new $^{60}\mbox{Fe}$
reservoir.

The new terrestrial reservoir we propose are microfossils comprised
of single-domain crystals of magnetite ($\mbox{\mbox{Fe}}_{3}\mbox{O}_{4}$),
produced by magnetotactic bacteria extant with this SN event. Provided
these bacteria preferentially uptake iron from fine-grained iron oxides
and ferric hydroxides, these microfossils should contain a biogenic
signature of this SN event. We have characterized the iron mass concentration
within the microfossils of a Pacific Ocean deep-sea drill core contemporaneous
with the SN event. In contrast with a previous $^{60}\mbox{Fe}$ search
in a sediment drill core extracted from the Norwegian Sea that found
no detectable $^{60}\mbox{Fe}$ signal~\citep{fitoussi08}, we deduce
that these microfossils should contain $^{60}\mbox{Fe/\mbox{Fe}}$
concentrations within the sensitivity limits of AMS and, moreover,
that there is sufficient quantity of microfossils in our drill core
to perform the AMS measurements.

\section{Magnetofossil Reservoir\label{sec:reservoir}}

Magnetotactic bacteria are single cell eukaryotes that are are unique
in that they produce intracellular crystal chains of single-domain
(SD) magnetite ($\mbox{\mbox{Fe}}_{3}\mbox{O}_{4}$) crystals called
\emph{magnetosomes}~\citep{frankel_nat_rev}. Putative fossil magnetosomes
(magnetofossils) have been dated as far back as 1.9~Gyr before the
present~\citep{kopp}.

Supernova iron arriving in the Earth's atmosphere in atomic/molecular
form, or in the form of nanometer-sized oxides, is expected to rapidly
react in aqueous environments via dissolution, binding to organic
ligands, and re-precipitation in the form of poorly crystalline ferric
hydroxides (FHO)~\citep{Jickells05}. In this form, Fe has a very
short residence time of $<100$~yr~\citep{baar} in the global ocean
system and finally becomes incorporated into the sediment by settling,
or through chemical reactions involving the dissolved form such as
with ferromanganese crusts. Organic complexes, FHO, and iron oxide
nanoparticles incorporated into the sediment are easily dissolved
and reprecipitated through sedimentary redox reactions. Specialized
classes of bacteria actively drive the iron redox cycle by reducing
available Fe(III) to Fe(II), which is then incorporated into new minerals
such as magnetite $(\text{Fe}_{3}\text{O}_{4})$ and greigite $(\text{Fe}_{3}\text{S}_{4})$.
Among these are the dissimilatory metal reducing bacteria (DMRB),
which extracellularly induce the precipitation of secondary iron minerals
such as siderite $(\text{Fe}_{3}\text{CO}_{3})$ and magnetite~\citep{zachara02};
while magnetotactic bacteria intracellularly grow magnetite or greigite
crystals~\citep{faivre08}. From culture studies, such DMRB have
been shown to preferentially utilize fine-grained iron oxides and
ferric hydroxides as their iron source over the bulk-sized secondary
minerals~\citep{roden96,bosch10}. This implies that those particles
of high surface to volume ratios (i.e. the fine-grained nanometer-sized
grains) are preferentially uptaken by these bacteria. In particular,
surface normalized bacterial Fe-reducing reaction rates are $\approx1.5-2$
orders faster~\citep{bosch10} for nano-sized ferrihydrite particles
as compared to grains with sizes comparable to bulk detrital grains.
Assuming these findings also hold for magnetotactic bacteria, it is
plausible that Fe processed by such bacteria almost entirely comes
from the Fe pool of easily available forms, which would include the
fine SN $^{60}\mbox{Fe}$-bearing grains, with little addition of
stable Fe from in-situ weathering of Fe-bearing primary minerals.
Because isotopic fractionation effects are negligible, the isotopic
composition of this Fe pool would then be reflected in the Fe isotopic
composition of the magnetosomes. It is therefore reasonable to consider
that those bacteria extant during the exposure of the Earth to the
SN-ejecta took up $^{60}\text{Fe}$ and incorporated it into their
magnetosomes, thus recording the signature of this event in the biological
record of our planet.

\section{Discovery Potential of Supernova $^{60}$Fe in Fossil Magnetosomes}

Because AMS makes a measurement of an atom ratio, the discovery potential
of $^{60}\mbox{Fe}$ in fossil magnetosomes hinges on not only the
incident flux of $^{60}\mbox{Fe}$ arriving on Earth, but also on
the degree to which the $^{60}\mbox{Fe}$ is diluted in sedimentary
marine reservoirs by influxes of stable Fe from the terrestrial iron-cycle.
We have seen that $^{60}\mbox{Fe}$ can be incorporated into secondary
minerals produced by DMRB or abiogenically during sediment diagenesis,
such as with ferromanganese crusts. In like manner, terrestrial Fe
is rendered to the sediment by these same mechanisms and, additionally,
by way of minerals sufficiently large and chemically resistant to
remain unaltered and, thus, survive diagenesis. Iron in primary minerals
is exclusively of terrestrial origin, while all secondary minerals,
whose formation was coeval with the SN event, will contain both $^{60}\mbox{Fe}$
and terrestrial Fe. Thus, those sediments having had conditions conducive
for the support of magnetotactic bacteria and the Fe redox cycle,
while having minimal detrital inputs, are best suited for detecting
a SN event. 

As a demonstrative example we discuss the case of an Ocean Drilling
Project sediment core (ODP core 848, Leg 138), from the Eastern Equatorial
Pacific $(2^{\circ}59.6'\,\text{S},\,110^{\circ}29'\,\text{W}$, 3.87~km
water depth), focusing on the 2.4 -- 3.3~Myr age interval corresponding
to the SN event reported in \citet{knie04}. The sediment is a pelagic
carbonate (60 -- 80\% $\text{CaCO}_{3}$, 20 -- 30\% $\text{SiO}_{2}$)
with a total iron content of 1.5 -- 3.5~wt\%~\citep{billeaud}.
The core-depth age correlation was established by comparison of the
Earth's magnetic field direction, as recorded by magnetic minerals
in the sediment, with a reference polarity time scale~\citep{shackleton95}.
The mean sedimentation rate of core 848 for the 2.4 -- 3.3~Myr age
interval is 0.54~cm/kyr~\citep{pisias95}. The location of this
core, being far removed from continental landmasses, should minimize
detrital Fe inputs from coastal runoff.

\begin{figure}[h]
\begin{centering}
\includegraphics[scale=0.49]{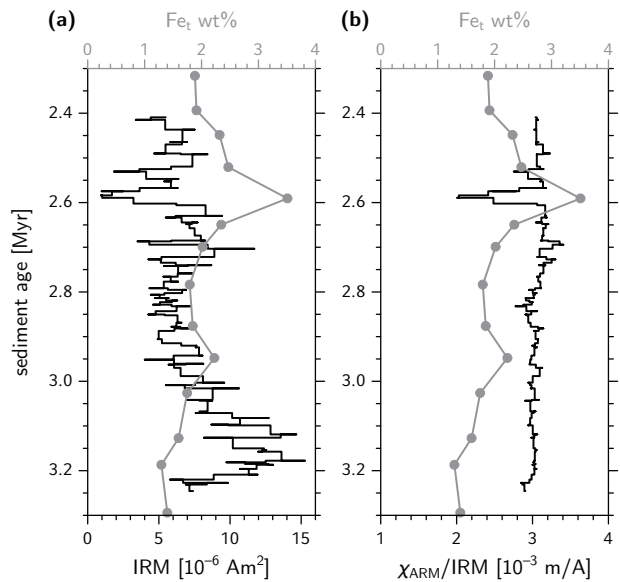} 
\par\end{centering}

\caption{IRM and $\chi_{\text{ARM}}/\text{IRM}$ results from $\approx5$~g
samples as a function of sediment age in ODP drill core 848, Leg 138.
Panel (a) shows our IRM results, while panel (b) shows our $\chi_{\text{ARM}}/\text{IRM}$
results. The grey data points display the known iron percentage and
are to be read from the top horizontal axis.\label{fig:irm}}
\end{figure}

The concentration of magnetofossils in this sediment core can be estimated
by measuring two types of remanent magnetizations acquired in the
laboratory: a so-called isothermal remanent magnetization (IRM), acquired
in a 0.1~T field; and an anhysteretic remanent magnetization (ARM),
acquired in a slowly decaying alternating field with an initial amplitude
of 0.1~T superimposed on a 0.1~mT field of constant bias. The magnetic
moments imparted by both magnetizations to $\approx5\,\mbox{g}$ samples
of powdered sediment were measured in a 2G superconducting rock magnetometer.
The IRM measurements are shown in panel (a) of Fig.~\ref{fig:irm}
as a function of sediment age. While an IRM records the remanent magnetization
of all ferrimagnetic minerals in the sample, regardless of their domain
state, ARM is predominantly acquired by single-domain (SD) ferrimagnetic
grains, or linear chains of such grains, that are well dispersed in
a non-magnetic matrix~\citep{egli02}; therefore, the ratio between
ARM susceptibility $\chi_{\text{ARM}}$ (ARM normalized by the bias
field) to IRM, $\chi_{\text{ARM}}/\text{IRM}$, is a sensitive domain
state indicator that can be used to discriminate between primary and
secondary ferrimagnetic minerals based on the large difference in
typical grain sizes between the two types. Panel (b) of Fig.~\ref{fig:irm}
shows our measured $\chi_{\text{ARM}}/\text{IRM}$ values as a function
of sediment age. Well dispersed SD particles are characterized by
$\chi_{\text{ARM}}/\text{IRM}>10^{-3}\,\mbox{m/A}$, while for most
primary minerals $\chi_{\text{ARM}}/\text{IRM}<2\times10^{-4}\,\mbox{m/A}$.
The SD particles produced by DMRB, which themselves would also contain
$^{60}\mbox{Fe}$, are expected to yield $\chi_{\text{ARM}}/\text{IRM}\approx2\times10^{-4}\,\mbox{m/A}$~\citep{moskowitz93}
provided the value measured on a cell culture~\citep{moskowitz93}
can be extrapolated to natural conditions. Intact magnetosome chains
are known, however, to have the highest reported values of $3_{-1}^{+2}\times10^{-3}\,\mbox{m/A}$~\citep{egli04,moskowitz93}.
Our drill core sediment has values, shown in panel b) of Fig.~\ref{fig:irm},
consistent with that expected for intact magnetosomes. We therefore
conclude that the IRM of this core section is almost entirely carried
by secondary SD minerals with a high proportion of magnetofossils.
Further evidence for the predominance of magnetofossils as the primary
magnetic remanence carriers in this core section is also provided
by more specific measurements (not shown here) which fulfill the detection
criteria stated in \citet{egli10}%

These IRM data in combination with the estimated supernova $^{60}\mbox{Fe}$
fluence determined in \citet{knie04} permit an estimate of the $^{60}\mbox{Fe}/\text{Fe}$
ratio in this core section under the reasonable assumption that the
measurements shown in Fig.~\ref{fig:irm} are representative of well
dispersed SD magnetite. Since the switching fields of secondary ferrimagnetic
minerals are smaller than the 0.1~T field employed to magnetize our
sediment samples, the IRM corresponds to a saturation remanent magnetization,
which, for non-interacting, uniaxial SD particles, corresponds to
half the saturation magnetization.

The iron concentration $\mathcal{C}_{\text{SDFe}}$ contained in SD
magnetite from this core can be estimated by way of,\begin{equation}
\mathcal{C}_{\text{SDFe}}=\frac{2m_{\text{rs}}}{w\mu_{s}}\frac{M_{\text{Fe}}}{M_{\text{Fe}_{3}\text{O}_{4}}}\label{eq:cfe}\end{equation}
where $m_{\text{rs}}$ and $\mu_{s}$ are the IRM-determined magnetic
moment and saturation magnetization, respectively, of the magnetic
mineral, $w$ is the IRM sample mass, and $M$ is the atomic mass
of iron or magnetite as implied by the corresponding subscript label.
From panel (a) of Fig.~\ref{fig:irm}, an average value of $m_{\text{rs}}\approx6\times10^{-6}\,\text{Am}^{2}$
is taken over the entire age span. The saturation magnetization of
magnetite is $\mu_{s}=92\,\text{Am}^{2}/\text{kg}$, and the mass
of each measured core sample was $w\approx5\,\text{g}$. These numbers
yield an estimated single domain iron mass concentration of $\mathcal{C}_{\text{SDFe}}\approx1.9\times10^{-5}\,\text{g/g}$.

The corresponding flux of SD Fe in this core sample is determined
by\begin{equation}
\Phi_{\text{Fe}}=\mathcal{C_{\text{SDFe}}}\frac{N_{A}}{M_{\text{Fe}}}\rho\frac{dh}{dt}\label{eq:flux_iron}\end{equation}
where $N_{A}$ is Avogadro's constant, while $\rho$ and $dh/dt$
are, respectively, the estimated density and sedimentation rate of
the core material. We take $\rho\approx2.7\,\mbox{g/cm}^{3}$, consistent
with that of calcium carbonate, along with the known sedimentation
rate~\citep{pisias95} $dh/dt=0.54\ \mbox{cm/kyr}$. With these numbers,
Eq.~\ref{eq:flux_iron} yields $\Phi_{\text{Fe}}\approx2.9\times10^{17}\,\text{atom/cm}^{2}\text{kyr}$
in SD magnetite.

From the work of Ref.~\citet{knie04}, the estimated $^{60}\mbox{Fe}$
terrestrial fluence, $\phi_{60}$, after correcting for the new $^{60}\mbox{Fe}$
half life, is given by,\begin{equation}
\phi_{60}=\frac{2^{\left(1/t_{1/2}-1/t_{1/2}^{\prime}\right)\, t}}{U_{\text{Fe}}}\phi_{60}^{\prime}=2.8\times10^{8}\ \mbox{atom}\mbox{/cm}^{2}\,,\label{eq:60fluence}\end{equation}
where $t=2.8$~Myr is the nominal time of the SN event, $t_{1/2}^{\prime}=1.5$~Myr
is the old $^{60}\mbox{Fe}$ half life, $t_{1/2}=2.62$~Myr is the
revised $^{60}\mbox{Fe}$ half life~\citep{rugel09}, $U_{\text{Fe}}=0.6\%$
is the Fe-uptake factor~\citep{knie04} of the FMC, and $\phi_{60}^{\prime}=2.9\times10^{6}\ \mbox{atom}\mbox{/cm}^{2}$
is the FMC $^{60}\mbox{Fe}$ fluence measured by~\citet{knie04}.
The width of the $^{60}\text{Fe/Fe}$ concentration spike of Fig.~1
in \citet{knie04} suggests a characteristic exposure time, $\tau$,
of Earth to this fluence of $\tau\approx500\,\text{kyr}$, assuming
that the residence time of iron in the ocean $(<100\,\text{yr})$
\citep{baar} is short compared to the exposure time. With these,
the terrestrial flux of supernova $^{60}\mbox{Fe}$ is estimated using\begin{equation}
\Phi_{60}=\frac{\phi_{60}}{\tau}\,.\label{eq:sn_flux}\end{equation}
Although we have adopted the width of the spike as the characteristic
exposure time, it is possible that this value could be much shorter
$(\tau\approx10\,\text{\text{kyr}})$~\citep{fields05}. However,
as Eq.~\ref{eq:sn_flux} shows, a shorter exposure time can only
serve to increase the value of our flux; we therefore use the nominal
value of $\tau\approx500\,\text{kyr}$ for our prospect estimates.

Under the assumption stated in Section~\ref{sec:reservoir} that
the bacteria will preferentially utilize the fine-grained iron-oxides
and ferric hydroxides as their source of iron from which they construct
their magnetosomes, an estimate of the resulting $^{60}\text{Fe/Fe}$
ratios contained in the magnetofossils is obtained by taking the ratio
of Eq.~\ref{eq:sn_flux} with Eq.~\ref{eq:flux_iron}.

Table~\ref{tab:flux} shows the resulting estimated $^{60}\text{Fe/Fe}$
ratios for three values of exposure time $\tau$ (250, 500 and 750~kyr)
and for two different terrestrial supernova $^{60}\mbox{Fe}$ fluences:
$\phi_{60}=2.8\times10^{8}\,\text{atom/cm}^{2}$ as described above
in Eq.~\ref{eq:60fluence}, and the more conservative case of $U_{\mbox{Fe}}\times\phi_{60}$.
This conservative case is also considered because the FMC Fe-uptake
factor is presently being reevaluated~\citep{korschinek_priv} and
it may, in fact, be unity. When considering that $^{60}\mbox{Fe}/\mbox{Fe}\approx2\times10^{-15}$
in the FMC $^{60}\mbox{Fe}$ discovery~\citep{knie04}, Table~\ref{tab:flux}
shows that, in both fluence cases, the resulting $^{60}\text{Fe}/\text{Fe}$
ratios are well within the capabilities of the Maier-Leibnitz AMS
facility across all exposure times $\tau$.%
\begin{table}[t]
\centering{}\begin{tabular}{r|c|c|c|}
\cline{2-4} 
 & \multicolumn{3}{|c|}{$^{60}$Fe/Fe}\tabularnewline
\cline{2-4} 
 & $\tau=250$ kyr  & $\tau=500$ kyr  & $\tau=750$ kyr \tabularnewline
\hline
\multicolumn{1}{|c||}{$\phi_{60}$} & $1.8\times10^{-12}$  & $9.0\times10^{-13}$  & $6.0\times10^{-13}$ \tabularnewline
\multicolumn{1}{|c||}{$U_{\text{Fe}}\times\phi_{60}$} & $1.1\times10^{-14}$  & $5.4\times10^{-15}$  & $3.6\times10^{-15}$ \tabularnewline
\hline
\end{tabular}\caption{Estimated $^{60}$Fe/Fe ratios in single domain magnetite in Site
848 drill core for different SN-ejecta exposure times and two $^{60}\mbox{Fe}$
fluences.\label{tab:flux}}
\end{table}

Remaining is the issue of the quantity of core material that would
be required to achieve a meaningful result from an AMS measurement.
The number of $^{60}\text{Fe}$ atoms per gram of drill core material
is given by,\begin{equation}
N_{60}=\mathcal{C}_{\text{SDFe}}\,\frac{N_{A}}{M_{\text{Fe}}}\left(\frac{^{60}\text{Fe}}{\text{Fe}}\right)\label{eq:n60}\end{equation}
where all symbols have been previously defined. Inspecting Table~\ref{tab:flux},
let us take $^{60}\text{Fe}/\text{Fe}\approx3.6\times10^{-15}$ as
our most conservative estimate of this ratio. We then obtain $N_{60}\approx760/\mbox{g}$
of drill core material. The total efficiency $\epsilon$ of the AMS
facility for $^{60}\text{Fe}$ is known to be $\epsilon\approx10^{-4}$.
Thus, per gram of drill core material, the $^{60}\text{Fe}$ counting
estimate should be something on the order of $\epsilon\cdot N_{60}\approx7.6\times10^{-2}$
atoms. A modest $\approx200$~g of sediment material, yielding $\approx3.8$~mg
of single-domain iron, would be sufficient for an AMS measurement
in search of this signal. Based on these estimates, we conclude there
is a plausible prospect for detecting a biogenic signature of this
supernova event in magnetofossils.

\section{Conclusion}

Radioactive $^{60}\text{Fe}$ is a definitive proxy for supernovae
and it has been previously discovered in a Pacific Ocean deep-sea
FMC~\citep{knie04}. Atmospheric ferric hydroxide nano-grains bearing
$^{60}\mbox{Fe}$ and settling to the deep-ocean should be rapidly
consumed by magnetotactic bacteria, resulting in the incorporation
of $^{60}\mbox{Fe}$ into their magnetosomes. Our magnetic remanence
measurements of ODP drill core 848 show that magnetofossils dominate
the magnetic signature of that portion of the drill core coeval with
the SN time window. Using the half-life corrected $^{60}\mbox{Fe}$
flux estimates determined in the FMC work of \citet{knie04} along
with the total magnetofossil Fe mass fraction determined from our
magnetic remanence results, we have estimated the $^{60}\mbox{Fe}/\mbox{Fe}$
ratio contained within the magnetofossils of this drill core. Our
most pessimistic estimate, $^{60}\mbox{Fe}/\mbox{Fe}\approx3.6\times10^{-15}$,
sits comfortably within the sensitivity capability of the existing
AMS facility at the Maier-Leibnitz laboratory. Remaining is an ambitious
experimental program dedicated to developing chemical and/or magnetic
extraction methods for the iron contained in the SD magnetic fraction
of this drill core segment, along with the concomitant AMS measurements
to determine if this iron contains supernova $^{60}\mbox{Fe}$. Efforts
on this front are underway.

\section*{Acknowledgements}

This work was supported by the DFG Cluster of Excellence \textquoteleft{}\textquoteleft{}Origin
and Structure of the Universe\textquotedblright{}. We want to acknowledge
the work of the following people: undergraduate students Christina
Loose and Raphael Kleindienst of TUM and Denise Schmidt of LMU, who
did much of the drill core sample preparation and magnetic measurements;
and LMU graduate student, Amy Chen, whose assistance was instrumental
in selecting a suitable ODP core. We further acknowledge the Ocean
Drilling Program for kindly providing us the drill core samples. Finally,
S. Bishop appreciates the proof reading efforts of Thomas Faestermann.

\bibliographystyle{elsarticle-num-names}
\bibliography{icarus_bib}

\end{document}